%% file: main.tex
\def\year{2022}\relax
\documentclass[letterpaper]{article} 
\usepackage{aaai22}  
\usepackage{times}  
\usepackage{helvet}  
\usepackage{courier}  
\usepackage[hyphens]{url}  
\usepackage{graphicx} 
\usepackage{natbib}  
\usepackage{caption} 
\DeclareCaptionStyle{ruled}{labelfont=normalfont,labelsep=colon,strut=off} 
\usepackage{algorithm}

\usepackage{eurosym}
\usepackage{subcaption}
\usepackage{makecell}
\usepackage{booktabs} 
\usepackage{multicol}
\usepackage{changepage}
\usepackage{csvsimple,longtable}
\usepackage[utf8]{inputenc}
\usepackage{adjustbox}
\usepackage{multirow}
\usepackage{caption}
\usepackage{geometry}
\usepackage{tcolorbox}
\usepackage{float}
\usepackage{changepage}
\usepackage{xcolor}
\usepackage[nottoc]{tocbibind}
\usepackage{fontawesome5}
\usepackage{subcaption}
\usepackage{tkz-euclide}
\usepackage{nameref}
\usepackage{glossaries}
\usepackage{makecell}
\usepackage{rotating}
\usepackage{chngcntr}

\usepackage{pgf}
\usepackage{url}
\usepackage{hyperref}
\usepackage{algorithm}
\usepackage{algpseudocode}
\usepackage{amsmath}
\usepackage{mdframed}
\usepackage{amsfonts}
\usepackage{tikz}
\usepackage{tikz-network}
\usetikzlibrary{positioning}

\tikzstyle{process} = [
    rectangle,
    minimum width=1cm, minimum height=1cm,
    text width=2cm,
    draw=black!0, fill=orange!20,
    rounded corners,
    align=center
]

\tikzstyle{arrow} = [thick,->,>=stealth]

\SetVertexStyle[Shape=circle, InnerSep=1, MinSize=3, FillColor=black, LineColor=black]
\SetEdgeStyle[Color=gray, Arrow=-stealth]

\usepackage{xcolor}

\newcommand{\answerTODO}[1]{\textcolor{red}{#1}} 
\usepackage{newfloat}
\usepackage{listings}
\floatstyle{ruled}
\newfloat{listing}{tb}{lst}{}
\floatname{listing}{Listing}
\title{TikTok StitchGraph: Characterizing communication patterns on TikTok through a collection of interaction networks}
\author{Mads H{\o}genhaug, Marcus Friis, Morten Pedersen, Luca Rossi\textsuperscript{*}
    }
\affiliations{
    IT University of Copenhagen, Denmark \\
    \textsuperscript{*}lucr@itu.dk
    
}

\begin{document}

\maketitle

\begin{abstract}
We present TikTok StitchGraph: a collection of $36$ graphs based on TikTok stitches. With its rapid growth and widespread popularity, TikTok presents a compelling platform for study, yet given its video-first nature the network structure of the conversations that it hosts remains largely unexplored. Leveraging its recently released APIs, in combination with web scraping, we construct graphs detailing stitch relations from both a video- and user-centric perspective. Specifically, we focus on user multi-digraphs, with vertices representing users and edges representing directed stitch relations. From the user graphs, we characterize common communication patterns of the stitch using frequent subgraph mining, finding a preference for stars and star-like structures, an aversion towards cyclic structures, and directional disposition favoring in- and out-stars over mixed-direction structures. These structures are augmented with sentiment labels in the form of edge attributes. We then use these subgraphs for graph-level embeddings together with Graph2Vec, we show no clear distinction between topologies for different hashtag topic categories. Lastly, we compare our StitchGraphs to Twitter reply networks and show that a remakable similarity between the conversation networks on the two platforms.  
\end{abstract}

\section{Introduction}

In the 21st century, the landscape of public discourse has undergone a significant transformation with the rise of short-form video content platforms such as TikTok, Instagram Reels, and YouTube Shorts.  Among these, TikTok has seen a significant increase in popularity, becoming one of the most popular social media services, with more than $1.67$ billion active monthly users worldwide \citep{tiktok_popular}. In particular, teens have heavily adopted this new format type \citep{tiktok_teens}. With the recent introduction of the TikTok Research API \citep{tiktokresearchapi}, performing large scale studies of TikTok has become a possibility. 

The short-form video format has been adopted for all communicative purposes, with content ranging from comedic sketches to serious political debates. This multifaceted nature of TikTok’s content makes it a compelling platform to study. The shift from traditional text-dominated social media, such as Facebook and X (formerly known as Twitter), to short-form video content presents new challenges in computationally analyzing public discourse. While natural language processing methods have sufficed for older platforms, the multimodal character of short-form video content introduces analytical challenges that require a combination of visual, auditory, and textual analysis.

This paper presents TikTok StitchGraph - a network of stitches - : possibly the first graph dataset detailing stitches on TikTok. A stitch network is constructed comprising $36$ different hashtags, containing all stitches created in May $2024$ using one of said hashtags. The dataset aims to explore and improve understanding of how people communicate using stitches on TikTok. The topological structure of TikTok stitch networks and the content of individual videos are analyzed to uncover patterns and insights into this emerging form of public discourse. Additionally, these findings are compared to Twitter to highlight structural differences and commonalities in how users interact across these two social media platforms. The approach combines basic network analysis, sentiment analysis, frequent subgraph mining, and network embeddings to gain an introductory understanding of the stitch networks collected from TikTok. Specifically, this paper aims to answer the following research questions:

\begin{itemize}
    \item How can stitch communication on TikTok be studied through the lens of network analysis?
    \item What stitch patterns are characteristic of TikTok?
    \item How do the stitch patterns vary between different content themes on TikTok?
    \item How do the discovered properties of TikTok stitch networks compare with Twitter reply networks?
\end{itemize}

\subsection{TikTok and stitches}
TikTok users consume the platfomr through a personalized video feed, curated by TikTok's algorithm. TikTok presents videos to users, and based on user engagement, it adapts the recommended content. Users can interact with the content they see in a variety of ways: liking it, commenting on it, re-posting it or via more complex interactions. Among the most interesting forms there are those based on the remix of the original video content: \textit{duets} and \textit{stitches}. \textit{Duets} enable users to create collaborative videos in a side-by-side or picture-in-picture format, facilitating interaction through real-time reactions or joint performances. 
\textit{Stitches}, introduced in in $2020$, allows users to create a "reaction" videos in response to another video, incorporating up to five seconds of the original -\textit{stitched} - video. This content remixing functionality yields an explicit network structure, where content can have direct connections to other content. This is akin to X's \textit{reply} and \textit{quote} functionality, where posts can be explicit reactions to other posts.  
Importantly, and differently from what happens on other platforms, users cannot stitch a video that is already a stitch of another video. Moreover, TikTok also gives users control over this feature, allowing them to disable stitching for specific videos or for all of their content. 

\section{Data and Materials}

To create a graph of TikTok stitches we started by using TikTok's API to collect all the videos posted with one of $36$ selected hashtags that have been selected to cover a diverse set of issues \footnote{The hashtags were chosen based on three criteria: comparable size, topic/community focus, and predominantly English language content. These are guiding criteria, but were not strongly enforced in our selection. The size criterion is there to limit the impact of size differences on the analysis of TikTok graphs. The topic/community focus criterion means that the hashtag should be centered on a topic, event, or community of people. The intent is to find hashtags where a conversation occurs such that users communicate using stitches. Lastly, the English language criterion guarantees that we can work with the auditory component of the videos, facilitating NLP analysis. The hashtags are: \#comedy, \#booktok, \#anime, \#storytime, \#lgbt, \#palestine, \#gaming, \#maga, \#football, \#catsoftiktok, \#news, \#trump2024, \#kpop, \#makeup, \#gaza, \#dogsoftiktok, \#gym, \#israel, \#learnontiktok, \#movie, \#challenge, \#blacklivesmatter, \#science, \#conspiracy, \#election, \#watermelon, \#biden2024, \#asmr, \#minecraft, \#prochoice, \#tiktoknews, \#plantsoftiktok, \#abortion, \#climatechange, \#jazz, \#guncontrol }. The data collection focused on videos created in May $2024$. Since the API does not provide direct support for identifying stitches, we leverage TikTok's default behavior of automatically including "\#stitch with @\textit{username}" with an hyperlink to the source video. This allowed us to scrape the original \textit{stitched} video. 

Once the stitch relationships are collected, metadata must be collected via TikTok API. At the end of this process, we have constructed a graph structure with the edge list scraped from TikTok and vertex metadata gathered through the TikTok API.
The result is a dataset composed of $36$ different graphs, with video metadata and audio data. While many many forms of data augmentation could be done we worked purely with speech content. Specifically, we extracted the sentiment of each video, adding an extra dimension to the graph data. Sentiment analysis is chosen because it captures the emotional tone of speech, offering a meaningful feature that aligns with our goal of enriching the graph without overcomplicating the task. 
To this end, we first use the Python library \textit{scenedetect} \citep{PySceneDetect}, which detects scene changes by analyzing variations in the HSL (Hue, Saturation, Lightness) color space to detect the scene shift that transitions from the stitchee part to the stitch part. We do this for the first 5 second of the video content since TikTok allow for up to 5 seconds of video to be stitched. Then we identify the language of each video with OpenAI's Whisper Model \citep{radford2022robustspeechrecognitionlargescale} and we apply Google's \textit{MediaPipe Audio Classifier} \citep{mediapipe_audio_classifier} to classify the audio type for each second of a video. This dimension provides insight into the dominant audio type within a video. Finally we use the Whisper base model ($74$ million parameters) \citep{radford2022robustspeechrecognitionlargescale} for transcriptions and We use \textit{VADER Sentiment Analysis} (hereafter referred to as VADER) \citep{VADER} to classifying the sentiment of the trascribed text into positive, neutral, or negative categories. When videos are unavailable for download, or when no transcription exists, they are labeled with a separate 'no content' label. 

On a general level the data we have collected can be used in two ways; as \textbf{a) video graphs} and \textbf{b) user graphs}. 

\textbf{a) Video graphs} are digraphs $G_v=(V,E)$, where vertices are TikTok videos, and edges are directed stitch relations. If video $u$ (the stitcher) stitches video $v$ (the stitchee), there is a directed edge $\{u, v\}$. The affordances and constraints of TikTok dictate the shape that video graphs can take. All video graphs consist of one or multiple stars $S_k$ ranging in size between a dyad $S_1$, and full star graph $S_{|E|}$, with the directionality always pointing towards the central vertex in the stars.

\textbf{b) User graphs} are multi-digraphs $G_u=(V,E)$, where vertices are users, and edges are stitch relations, but between users who stitch each other's videos. If user $u$ stitches a video from user $v$, then there is a directed edge $\{u,v\}$. Since users can create multiple stitches and be stitched multiple times, the topology of user graphs is not as rigid as that of video graphs. It allows for any arbitrary shapes along with self loops, since users can stitch videos from themselves, and multi-edges, since users can create multiple stitches of a video or different videos from another user. The size of the user graph is constrained by its respective video graph. The number of vertices in the user graph $|V_u|$ ranges from $1$ to the number of vertices in the video graph $|V_v|$, depending on the number of unique video creators. The number of edges for both the user- and video graph will always be equivalent $|E_u| = |E_v|$.

\begin{table}[ht] 
    \centering 
    \resizebox{\columnwidth}{!}
        { \input{figures/hashtag_table} } 
        \caption{Categorization of each collected hashtag. Each of the $36$ collected hashtags are assigned one of three categories: \textit{Shared Interest}, \textit{Entertainment}, or \textit{Political}. The category assignment is assigned manually based on the nature of the observed content.}
    \label{tab:hashtags} 
\end{table}

Each hashtag has been assigned to one of three categories (see \ref{tab:hashtags}): \textit{Shared Interest}, \textit{Entertainment}, or \textit{Political}. These categories are an attempt to reflect the overarching types of the hashtags and offer an additional dimension for comparing network topologies, contents, and metadata. Hashtags are selected to similarly cover all three categories. While some hashtags could fit into multiple categories, the assignment reflects the predominant use or context observed. For example, \textit{\#lgbt} could align with both Shared Interest and Political; however, watching videos from said hashtag showed that it is predominantly used to engage in a community of like-minded people instead of as a forum for political discussion, and for this reason is assigned to Shared Interest. Similarly, while \textit{\#comedy} might suggest a specific interest, it is categorized as Entertainment to reflect its broader context. The line between these categories can be ambiguous and the specific assignment of hashtags could arguably be changed, yet this framework allows for meaningful comparisons within and across groups. The hashtag \textit{\#watermelon} is particularly ambiguous, as it encompasses both occasional "viral food" recipes and association with the Palestinian flag.  Recently, this symbolism has resurfaced on TikTok as a form of algospeak, enabling users to discuss Palestine while avoiding algorithmic penalties. However, we find that during the time period in which the data was gathered, TikTok users predominantly used it to refer to actual watermelons. Therefore, it is categorized as Entertainment.

\subsection{Descriptive analysis}

Given the way in which TikTok's affordances shape the possible network of stitches, it is more interesting to focus on the \textit{user graphs}.
User graphs are constructed and presented in Table \ref{tab:user_table}. The number of edges between a video graph and its corresponding user graph is equivalent, due to user graphs being multi-digraphs, where each edge is a stitch between users. Although the user graphs have no structural constraints, they all display some of the same properties. Most notably, all of the graphs have essentially $0$ reciprocity and clustering. Furthermore, there is a large discrepancy between the undirected and directed diameters ($d_u$ and $d$), as well as the average undirected path length, $L_u$ and the average path length $L$ in the largest weakly connected component. Many of the largest components display high degree centralization, with $5$ of them achieving a score of $1$, meaning that they are stars. 

\begin{table*}[t]
    \include{figures/user_table_hybrid}
    \caption{Selected metrics for each of the $36$ collected user graphs, with metrics for both the full graphs and their largest weakly connected components, and with aggregate rows in the bottom for each category. Despite no topological constraints, user graphs all display no clustering or reciprocity, a significant difference between directed and undirected path lengths, and a high degree centralization in their largest component.}
    \label{tab:user_table}
\end{table*}

In addition to the metrics presented in the tables, we also noted some observations about the nature of stitches. As expected, stitchers generally have much fewer views than the stitchee, with the stitchee on average having $12$ thousand times as many views, showing tendencies of it being e.g. regular users stitching popular clips and adding their perspective or reaction. Similarly, the follower count of the stitchees is approximately $38$ times higher than that for the stitchers\footnote{For practical API rate limit reasons, this calculation is based on a subset of hashtags, namely: \textit{\#blacklivesmatter}, \textit{\#climatechange}, \textit{\#election}, \textit{\#jazz}, \textit{\#kpop}, \textit{\#maga}, \textit{\#movie}, \textit{\#science}, and \textit{\#tiktoknews}.}, and the total user-likes is approximately $40$ times greater. We also find that only $18\%$ of the stitched videos use the same hashtag as the stitcher, based on our scraped data, indicating that stitches are not confined to a single "community". A notable example of a stitchee exhibiting multiple of these properties is a video by a semi-popular user. It begins with the words, 'What do other girls have on the walls in their bedroom?' and includes only the hashtag \textit{\#greenscreen}, due to the use of TikTok's official greenscreen effect. The video is stitched $486$ times across $19$ hashtags, with stitchers responding by showcasing their bedroom walls usually decorated with a poster or something similar relating to the used hashtag, despite no community affiliation from the stitchee.

\subsection{Comparable Twitter data}

To provide a point of comparison we use Twitter data previously used in other research. Unlike TikTok’s video-based, multimodal content, Twitter’s reply networks are focused on text-based interactions, where users respond directly to tweets and other replies. For this paper, we use existing Twitter data, provided by \cite{doi:10.1126/sciadv.abq2044}. This data, from which we construct the reply networks, centers around six topics and events: gun control, pro-choice, abortion, a vice presidential debate, a second presidential debate, and the U.S. Supreme Court ruling on Obamacare. While comparable in some thematic way, the Twitter dataset is a collection of tweets containing at least one of a couple keywords per overall topic.

\section{Methods}

This paper examines the structural patterns of TikTok stitch networks to quantitatively compare the topologies of different graphs and hashtag categories in TikTok StitchGraph. By analyzing whether hashtags within the same category exhibit structural similarity, the goal is to uncover patterns in how users interact and communicate on the platform. Two main approaches are employed: subgraph analysis and graph representation learning. Subgraph analysis identifies recurring subgraphs that form the observed communication patterns within stitch networks, while graph representation learning focuses on deriving graph-level embeddings to represent the networks in a lower-dimensional space, enabling the evaluation of similarities and differences between hashtag groups within the learned vector representation. By combining subgraph analysis with graph embeddings, both micro-level patterns (e.g., motifs like star structures or chains) and macro-level patterns (e.g., clustering of hashtag categories) are detected. This dual approach facilitates a more comprehensive understanding of the communication dynamics on TikTok. 

\subsection{Frequent Subgraph Mining}
To identify characteristic stitch patterns on TikTok, the \textit{subgraphs} that constitute the observed TikTok stitch user graphs are examined. The objective is to identify the \textit{motifs} that characterize communicative patterns on TikTok. To this end, \textit{frequent subgraph mining} is employed, a technique used to discover subgraphs that appear repeatedly within a graph \citep{coscia2021atlasaspiringnetworkscientist}. Specifically, \textit{transactional graph mining} is applied, aiming to identify subgraphs that frequently occur across a collection of graphs. In practice, this involves pruning the subgraph search space by constraining the minimum required \textit{support} of a subgraph. In a transactional setting, the support of a subgraph is the number of graphs in the graph set with which the subgraph is \textit{isomorphic}. An important implication of the transactional setting is that the mined subgraphs may vary in how representative they are of specific graphs. Under the transactional support definition, a subgraph is treated as equally representative of all graphs in which it appears, even if its frequency within the graphs differs.
To identify subgraphs we used \textit{gSpan} \citep{yan2002gspan} and \textit{MoSS} \citep{borgelt2002moss} for undirected and directed structures, respectively. 
In mining undirected subgraphs, we prune the search tree by limiting the maximum substructure size to $|V|=4$ to prevent memory issues.

\subsection{Graph Embedding}
Graph embeddings provide a powerful framework to represent relationships and interactions in complex systems. By simplifying high-dimensional and intricate network structures, they map these systems into a lower-dimensional vector space, with the goal of capturing their structural and topological properties in numerical form.

In this study, graph embeddings are applied to analyze the structure of user graphs. Specifically, we embed and cluster $36$ distinct graphs, each representing a specific hashtag. We evaluate whether hashtags share structural properties, offering insight into how stitch patterns vary between content topics.

To encode the graphs, we use \textit{Graph2Vec} \citep{graph2vec}, implemented by the Karate Club library \citep{karateclub}. As a method for graph representation learning, Graph2Vec encodes entire graphs into fixed-length embeddings, preserving both structural and topological characteristics. Inspired by \textit{Word2Vec} \citep{word2vec} and \textit{Doc2Vec} \citep{doc2vec}, Graph2Vec represents individual graphs as "documents" and rooted subgraphs as "words." Using the Weisfeiler-Lehman relabeling strategy \citep{shervashidze2011weisfeiler}, it captures both local and global features of the graphs. It should be noted that Graph2Vec does not support edge-attributed graphs, and hence we refrain from embedding the sentiment graph with Graph2Vec. 

To address this limitation and to provide additional insights we adopt a Bag-Of-Subgraphs approach. The mined subgraphs from frequent subgraph mining can be used to define a subgraph-graph occurrence matrix to extract vector representations for entire graphs. This approach is analogous to the Bag-Of-Words model, but instead of using a vocabulary of words as basis vectors, it leverages subgraph isomorphisms. With this approach, each graph is represented as a collection of subgraph occurrences, ignoring their specific positions or arrangement within the graph. This yields, in contrast to Graph2Vec, an interpretable embedding space, where each dimension can be explained as the presence of a specific subgraph and can be applied to all variants of graphs.

The data collection was exstenvie and included some potentially sensitive elements (e.g., users ids, textual transcribe, etc). Nevertheless, since the research only focuses on netwrok structure all unecessary data has been deleted after the enriched stitch-graph has been constructed.

\section{Results}

\subsection{Subgraphs}

\begin{figure}[t]
\centering
\includegraphics[width=0.5\textwidth]{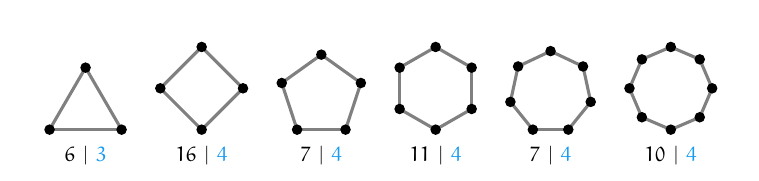} 
\caption{The figure shows the cyclic subgraphs identified within the largest connected components of user graphs, ranging from a triangle to an octagon. The numbers beneath denote the support of the subgraph in TikTok and Twitter in that order.}
\label{fig:cycle_subgraphs}
\end{figure}

The first notable observation is the lack of cyclic subgraphs. As seen in Figure \ref{fig:cycle_subgraphs}, the most common cycle is a square, with a support of $16$. Interestingly, the support of cycles with an even number of vertices is consistently higher than that of cycles with an odd number of vertices. An odd-numbered cycle means that a user has to be both a stitcher and a stitchee for these cycles to occur. This is indicative of it being rare for a user to both stitch and be stitched. In contrast, this tendency is not observed in the Twitter data. 

\begin{figure}[t]
\centering
\includegraphics[width=0.5\textwidth]{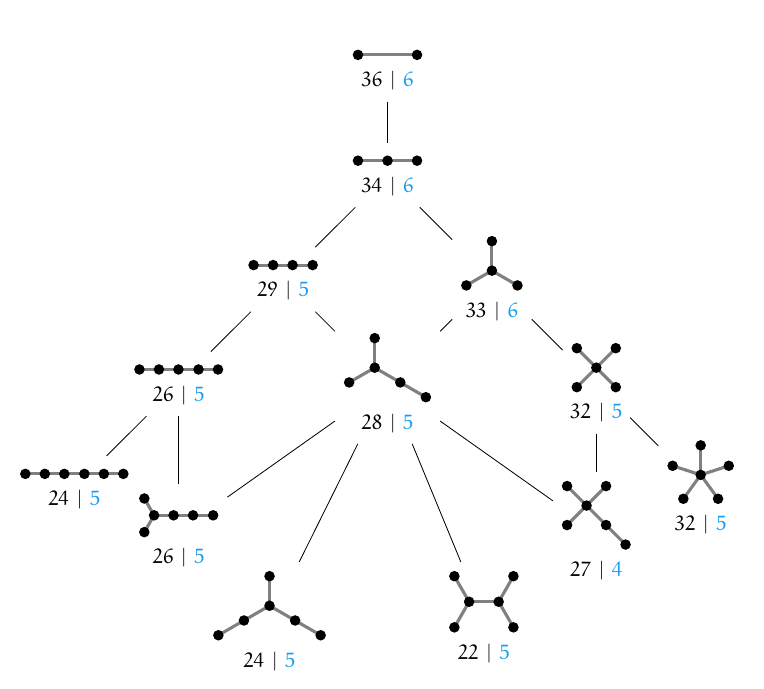} 
\caption{The figure shows the frequent undirected subgraphs identified in TikTok user graphs using gSpan on the largest connected component. The lines between subgraphs illustrate their hierarchical relationships, where each subgraph extends from the one above. The numbers are the supports of the given subgraphs in TikTok and Twitter respectively. Simpler structures, such as dyads $S_1$ ($support=36$), form the foundation, while more complex patterns like chains and star-like structures emerge as extensions with lower support. This highlights how TikTok stitch patterns evolve, often around central hubs or sequential interactions.}
\label{fig:subgraph_hierarchy}
\end{figure}

We analyze the complete subgraph hierarchy up to six vertices can be mapped out, as presented in Figure \ref{fig:subgraph_hierarchy}. From it, the hierarchy of subgraphs shows how the support of any child subgraph is at most equal to its parent. We see that stars and star-like patterns generally have slightly higher support than chains and chain-like patterns. Furthermore, we note that, without cycles, any mined subgraph will always be an interpolation between stars and chains.

\subsection{Sentiment Subgraphs}

\begin{figure}[t]
\centering
\includegraphics[width=0.5\textwidth]{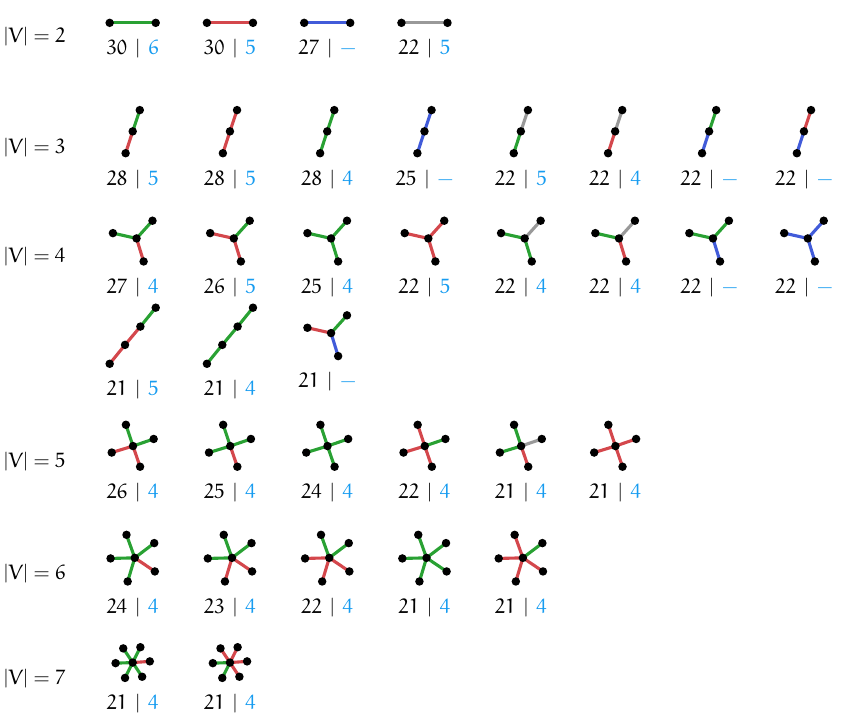} 
\caption{The figure shows the frequent undirected subgraphs identified in TikTok user graphs using gSpan on the largest connected component. The lines between subgraphs illustrate their hierarchical relationships, where each subgraph extends from the one above. The numbers are the supports of the given subgraphs in TikTok and Twitter respectively. Simpler structures, such as dyads $S_1$ ($support=36$), form the foundation, while more complex patterns like chains and star-like structures emerge as extensions with lower support. This highlights how TikTok stitch patterns evolve, often around central hubs or sequential interactions.}
\label{fig:sentiment_subgraphs}
\end{figure}

Adding sentiment as a dimension yields the subgraphs illustrated in Figure \ref{fig:sentiment_subgraphs}. From these we see that positive and negative edges have a support of $30$, missing sentiment edges $27$, and neutral edges $22$. 
A notable observation is the lack of anything between pure stars and chains, unlike in the previous subgraph mining. In fact, subgraphs with $|V| \geq 5$ consist of purely stars, further indicating that TikTok stitch networks are dominated by this pattern although no mechanism prevents other subgraphs from appearing. It is also interesting to observe how both \textit{positive} and \textit{negative} stars are present.

\subsection{Graph Embedding}

\begin{figure}[t]
\centering
\includegraphics[width=0.5\textwidth]{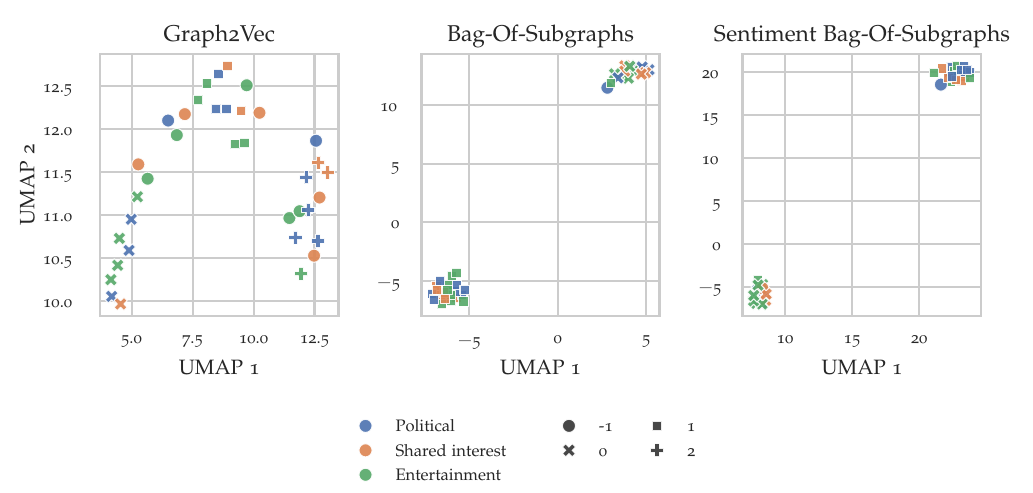} 
\caption{Comparison of user graphs across three embedding spaces, labeled by assigned hashtag categories, clustered using HDBSCAN, and subsequently dimensionality reduced with UMAP. They reveal no separation between graphs corresponding to different categories. This observation is further reinforced by the minimal overlap between the assigned categories and the clusters identified by HDBSCAN.}
\label{fig:emb_scatter}
\end{figure}

To compare the collected graphs with clustering, we employ various graph embedding techniques to find a representation that facilitates vectorized comparisons. To this end, we employ two approaches: we apply the graph representation learning algorithm Graph2Vec, and we use the identified subgraphs in a Bag-Of-Subgraphs approach, leading to the results seen in Figure \ref{fig:emb_scatter}.

Inspecting the figure reveals no clear separation between graphs of different categories in any embedding space, indicating no relationship between the graph topologies and the categorization in the embedding spaces. Moreover, in the Graph-2Vec representation, three clusters are detected, whereas both Bag-Of-Subgraphs reveal only two clusters. Investigating the driving mechanism behind this partitioning, we find that the size of the graphs almost entirely explain the clustering. In the Graph2Vec space, the average graph size for cluster $0$ is $6.3$, $1$ is $55.9$, and $2$ is $230.7$, and the same trend is observed in the other two representations, showing a strong positive correlation between cluster assignment and graph size. These detected clusters show no overlap with the categorization. Calculating Normalized Mutual Information (NMI) between the categorization and the detected clusters supports this observation, with NMI values ranging from $0.001$ to $0.054$, further confirming the lack of overlap.

\begin{figure}[t]
\centering
\includegraphics[width=0.5\textwidth]{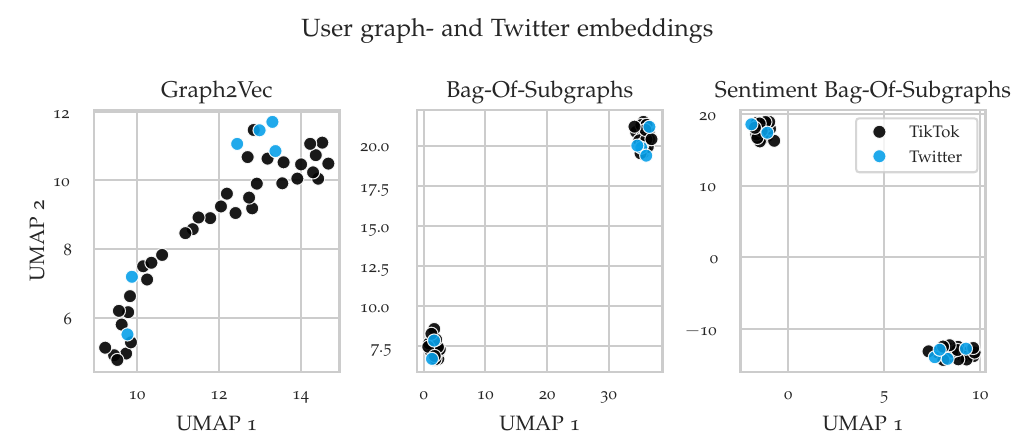} 
\caption{Comparison of user graph embeddings from TikTok and Twitter across three embedding spaces, dimensionality reduced with UMAP. While no distinct separation is observed for most graphs, larger Twitter graphs appear to occupy a slightly different region compared to TikTok graphs in the Graph2Vec embedding space. No such distinction is evident in either Bag-Of-Subgraphs.}
\label{fig:twitter_scatter}
\end{figure}
Similar to the previous results, Figure \ref{fig:twitter_scatter} reveals no clear distinction between TikTok user graphs and Twitter reply networks across both variants of Bag-Of-Subgraph. However, in the Graph2Vec embedding space, larger Twitter graphs occupy a distinct region compared to TikTok graphs, suggesting potential differences in graph structures for larger graphs. Consistent with observations in the TikTok-only analysis, graph size remains the primary factor driving separation across all embedding spaces. Smaller Twitter graphs tend to cluster near smaller TikTok user graphs, while larger graphs from both platforms are similarly grouped. Overall, the embeddings do not show strong platform-specific grouping for the majority of graphs.

\section{Conclusion}
In this work we presented the what is to our knowledge the first network study of TikTok conversational dynamics. We have collected the TikTok StitchGraph dataset: a starting point to study TikTok’s stitch-based interactions, covering $36$ hashtags across different topics. The graph structures, comprising both video and user graphs, generally exhibit sparse, star-like patterns with high degree centralization, short path lengths, low clustering, and low reciprocity. Together, they reflect how stitches on TikTok often center around key videos and/or users. This structure, that is observed across the diverse set of $36$ topical hashtags we have studied shows how the stitch feature in TikTok seems to be more used to gain views and attention by piggybacking other users' audience than to support true conversational dynamics. 
 
Using Graph2Vec and the Bag-Of-Subgraphs approach, user graphs are represented in vector format. However, doing so reveals no clear relationship between topic and topology, as both approaches primarily capture size-related features rather than topic-specific differences. 

Comparisons with Twitter reply networks reveal similar patterns highlighting how size-related factors dominate in the embeddings of both platforms. Larger Twitter graphs display minor separation from the larger user graphs in the Graph2Vec space, yet these are not captured using clustering. There is no separation in the Bag-Of-Subgraphs spaces, likely due to the subgraphs being mined from TikTok. Without using subgraphs distinct to Twitter, it might not capture unique patterns exclusive to Twitter. Further research should consider graph sizes to minimize its influence as the primary differentiator. However, given the differences in data collection, as well as in social use of the two platforms, comparative results should be approached with caution.

\bibliography{aaai22}

\subsection{Ethics Checklist}

\begin{enumerate}

\item For most authors...
\begin{enumerate}
    \item  Would answering this research question advance science without violating social contracts, such as violating privacy norms, perpetuating unfair profiling, exacerbating the socio-economic divide, or implying disrespect to societies or cultures?
    \answerTODO{Yes, user's privacy has been preserved and understanding if and how conversations and debates are structured by TikTok is relevant given the growing relevance of the platform.}
  \item Do your main claims in the abstract and introduction accurately reflect the paper's contributions and scope?
    \answerTODO{Yes}
   \item Do you clarify how the proposed methodological approach is appropriate for the claims made? 
    \answerTODO{We offered a detailed description of the methods and how they align with the RQs.}
   \item Do you clarify what are possible artifacts in the data used, given population-specific distributions?
    \answerTODO{While this doesn't fully apply to our case we discussed extensively possible artifacts and limitations emerging from data collection.}
  \item Did you describe the limitations of your work?
    \answerTODO{Yes}
  \item Did you discuss any potential negative societal impacts of your work?
    \answerTODO{We have considered this but given the nature of the work we can't really identify any clear negative impact.}
      \item Did you discuss any potential misuse of your work?
    \answerTODO{No potential misuse of understanding TikTok conversation network immediately comes to mind.}
    \item Did you describe steps taken to prevent or mitigate potential negative outcomes of the research, such as data and model documentation, data anonymization, responsible release, access control, and the reproducibility of findings?
    \answerTODO{Yes}
  \item Have you read the ethics review guidelines and ensured that your paper conforms to them?
    \answerTODO{Yes, to the best of our understanding.}
\end{enumerate}

\item Additionally, if your study involves hypotheses testing...
\begin{enumerate}
  \item Did you clearly state the assumptions underlying all theoretical results?
    \answerTODO{Not relevant}
  \item Have you provided justifications for all theoretical results?
    \answerTODO{Not relevant}
  \item Did you discuss competing hypotheses or theories that might challenge or complement your theoretical results?
    \answerTODO{Not relevant}
  \item Have you considered alternative mechanisms or explanations that might account for the same outcomes observed in your study?
    \answerTODO{Not relevant}
  \item Did you address potential biases or limitations in your theoretical framework?
    \answerTODO{Not relevant}
  \item Have you related your theoretical results to the existing literature in social science?
    \answerTODO{Not relevant}
  \item Did you discuss the implications of your theoretical results for policy, practice, or further research in the social science domain?
    \answerTODO{Not relevant}
\end{enumerate}

\item Additionally, if you are including theoretical proofs...
\begin{enumerate}
  \item Did you state the full set of assumptions of all theoretical results?
    \answerTODO{Not relevant}
	\item Did you include complete proofs of all theoretical results?
    \answerTODO{Not relevant}
\end{enumerate}

\item Additionally, if you ran machine learning experiments...
\begin{enumerate}
  \item Did you include the code, data, and instructions needed to reproduce the main experimental results (either in the supplemental material or as a URL)?
    \answerTODO{Not relevant}
  \item Did you specify all the training details (e.g., data splits, hyperparameters, how they were chosen)?
    \answerTODO{Not relevant}
     \item Did you report error bars (e.g., with respect to the random seed after running experiments multiple times)?
    \answerTODO{Not relevant}
	\item Did you include the total amount of compute and the type of resources used (e.g., type of GPUs, internal cluster, or cloud provider)?
    \answerTODO{Not relevant}
     \item Do you justify how the proposed evaluation is sufficient and appropriate to the claims made? 
    \answerTODO{Not relevant}
     \item Do you discuss what is ``the cost`` of misclassification and fault (in)tolerance?
    \answerTODO{Not relevant}
  
\end{enumerate}

\item Additionally, if you are using existing assets (e.g., code, data, models) or curating/releasing new assets...
\begin{enumerate}
  \item If your work uses existing assets, did you cite the creators?
    \answerTODO{Yes}
  \item Did you mention the license of the assets?
    \answerTODO{Not relevant}
  \item Did you include any new assets in the supplemental material or as a URL?
    \answerTODO{No}
  \item Did you discuss whether and how consent was obtained from people whose data you're using/curating?
    \answerTODO{Data was obtained through research API and following the DSA/GDPR transfer of data access for research activity.}
  \item Did you discuss whether the data you are using/curating contains personally identifiable information or offensive content?
    \answerTODO{Yes, personal information has been removed following a principle of data minimization}
\item If you are curating or releasing new datasets, did you discuss how you intend to make your datasets FAIR (see \citet{fair})?
\answerTODO{Not relevant}
\item If you are curating or releasing new datasets, did you create a Datasheet for the Dataset (see \citet{gebru2021datasheets})? 
\answerTODO{Not relevant}
\end{enumerate}

\item Additionally, if you used crowdsourcing or conducted research with human subjects...
\begin{enumerate}
  \item Did you include the full text of instructions given to participants and screenshots?
    \answerTODO{Not relevant}
  \item Did you describe any potential participant risks, with mentions of Institutional Review Board (IRB) approvals?
    \answerTODO{Not relevant}
  \item Did you include the estimated hourly wage paid to participants and the total amount spent on participant compensation?
    \answerTODO{Not relevant}
   \item Did you discuss how data is stored, shared, and deidentified?
   \answerTODO{Not relevant}
\end{enumerate}

\end{enumerate}

\end{document}

%% file: figures/hashtag_table.tex
\begin{tabular}{|l|l|l|}
    \hline
    \textbf{Shared Interest} & \textbf{Entertainment} & \textbf{Political} \\ \hline
    Anime & ASMR & Abortion \\ \hline
    Booktok & Catsoftiktok & Biden2024 \\ \hline
    Gaming & Challenge & Blacklivesmatter \\ \hline
    Gym & Comedy & Climatechange \\ \hline
    Jazz & Conspiracy & Election \\ \hline
    Kpop & Dogsoftiktok & Gaza \\ \hline
    Lgbt & Football & Guncontrol \\ \hline
    Makeup & Learnontiktok & Israel \\ \hline
    Minecraft & Movie & Maga \\ \hline
    Plantsoftiktok & News & Palestine \\ \hline
    & Science & Prochoice \\ \hline
    & Storytime & Trump2024 \\ \hline
    & Tiktoknews & \\ \hline
    & Watermelon & \\ \hline
\end{tabular}

%% file: figures/user_table_hybrid.tex
\begin{tabular}{l|rrrr|rrrrr}   \toprule   \multicolumn{1}{c|}{} & \multicolumn{4}{c|}{Full graph} & \multicolumn{4}{c}{Largest weakly connected component} \\             Hashtag &  $|V|$ &   \makecell{\#Compo-\\nents} & $D$ &  $D_u$ &  $|V|$ &     $C_u$ &  Reciprocity &  \makecell{Degree\\ centralization} \\    \midrule             comedy &   4608 &                          1135 &   2 &     19 &   1838 &       0.00 &         0.00 &                                0.03 \\            booktok &   3540 &                          1119 &   4 &     24 &    844 &       0.00 &         0.00 &                                0.09 \\          storytime &   2036 &                           762 &   2 &      7 &    173 &       0.00 &         0.00 &                                0.89 \\               lgbt &   1685 &                           566 &   2 &     14 &    365 &       0.00 &         0.00 &                                0.10 \\              anime &   1605 &                           443 &   4 &     16 &    646 &       0.00 &         0.01 &                                0.17 \\          palestine &   1236 &                           434 &   1 &     14 &    144 &       0.00 &         0.00 &                                0.22 \\       catsoftiktok &   1059 &                           354 &   2 &     11 &    142 &       0.00 &         0.00 &                                0.45 \\             gaming &   1023 &                           325 &   4 &     17 &    243 &       0.01 &         0.00 &                                0.13 \\           football &    935 &                           321 &   2 &     13 &     73 &       0.00 &         0.00 &                                0.27 \\       dogsoftiktok &    919 &                           385 &   1 &      5 &     13 &       0.00 &         0.00 &                                0.61 \\             makeup &    915 &                           345 &   2 &      5 &     68 &       0.00 &         0.00 &                                1.00 \\               kpop &    906 &                           236 &   3 &     13 &    344 &       0.00 &         0.00 &                                0.69 \\               gaza &    801 &                           269 &   2 &     11 &    109 &       0.00 &         0.00 &                                0.29 \\               news &    784 &                           255 &   2 &      6 &     73 &       0.00 &         0.00 &                                0.90 \\          trump2024 &    768 &                           202 &   2 &     14 &    272 &       0.00 &         0.00 &                                0.10 \\                gym &    742 &                           297 &   2 &      9 &     74 &       0.00 &         0.00 &                                0.48 \\               maga &    644 &                           106 &   3 &     12 &    414 &       0.00 &         0.00 &                                0.16 \\             israel &    594 &                           183 &   2 &     22 &    141 &       0.00 &         0.00 &                                0.07 \\              movie &    583 &                           199 &   3 &     12 &     71 &       0.03 &         0.00 &                                0.26 \\          challenge &    485 &                           191 &   2 &     10 &     35 &       0.00 &         0.00 &                                0.13 \\      learnontiktok &    464 &                           131 &   6 &      9 &     79 &       0.16 &         0.03 &                                0.15 \\            science &    411 &                           182 &   1 &      5 &     11 &       0.00 &         0.00 &                                0.51 \\   blacklivesmatter &    388 &                           119 &   2 &      8 &     64 &       0.00 &         0.00 &                                0.41 \\         conspiracy &    365 &                           150 &   2 &      4 &     11 &       0.00 &         0.00 &                                0.88 \\           election &    334 &                           131 &   1 &      6 &     11 &       0.00 &         0.00 &                                0.39 \\         watermelon &    290 &                           119 &   1 &      6 &     17 &       0.00 &         0.00 &                                0.29 \\               asmr &    191 &                            92 &   1 &      2 &      6 &       0.00 &         0.00 &                                1.00 \\          minecraft &    151 &                            66 &   1 &      3 &     14 &       0.00 &         0.00 &                                1.00 \\          biden2024 &    188 &                            50 &   2 &     11 &     63 &       0.00 &         0.00 &                                0.23 \\          prochoice &    141 &                            46 &   1 &      4 &     30 &       0.00 &         0.00 &                                0.67 \\         tiktoknews &    120 &                            49 &   1 &      2 &      8 &       0.00 &         0.00 &                                1.00 \\     plantsoftiktok &     96 &                            58 &   1 &      6 &      8 &       0.00 &         0.00 &                                0.24 \\           abortion &     87 &                            37 &   1 &      3 &      6 &       0.00 &         0.00 &                                1.00 \\      climatechange &     86 &                            43 &   1 &      3 &      4 &       0.00 &         0.00 &                                0.33 \\               jazz &     32 &                            16 &   1 &      1 &      2 &       0.00 &         0.00 &                                 - \\         guncontrol &     10 &                             5 &   1 &      1 &      2 &       0.00 &         0.00 &                                 - \\   \midrule   \textit{Shared interest} & 1069.5 & 347.1 & 2.4 & 10.8 & 260.8 & 0.00 & 0.00 & 0.43 \\     \textit{Entertainment} &  946.4 & 308.9 & 2.0 &  7.9 & 182.1 &  0.01 & 0.00 & 0.53 \\         \textit{Political} &  439.8 & 135.4 & 1.6 &  9.1 & 105 &   0.00 & 0.00 & 0.35 \\   \bottomrule\end{tabular}